\documentclass[acmtog,authorversion,nonacm]{acmart} 
\acmSubmissionID{488}

\usepackage{booktabs} 
\usepackage{enumitem}
\usepackage{amsmath}
\usepackage{bm}
\usepackage{svg}
\usepackage{sidecap}

\definecolor{peach}{rgb}{ 0.943, 0.188, 0.526}
\definecolor{plum}{rgb}{ 0.858, 0.188, 0.478}
\definecolor{muted_navy_blue}{RGB}{63, 75, 166}
\definecolor{muted_sky_blue}{RGB}{134,166,213}
\definecolor{federal_blue}{RGB}{0,96,240}
\definecolor{regulation_red}{RGB}{226, 20, 79}
\definecolor{federal_gold}{RGB}{240, 212, 14}
\definecolor{douggreen}{RGB}{39,131,48}

\def\myeqref#1{Eq.~\ref{#1}}

\usepackage[ruled]{algorithm2e} 

\SetAlFnt{\small}
\SetAlCapFnt{\small}
\SetAlCapNameFnt{\small}
\SetAlCapHSkip{0pt}

\acmJournal{TOG}

\begin{document}
\title{Dress Anyone : Automatic Physically-Based Garment Pattern Refitting}


\author{Hsiao-yu Chen}
\email{hsiaoyu@meta.com}
\affiliation{%
  \institution{Meta Reality Labs}
  \country{USA}
}
\author{Egor Larionov}
\affiliation{%
  \institution{Meta Reality Labs}
  \country{USA}
}
\author{Ladislav Kavan}
\affiliation{%
  \institution{Meta Reality Labs}
  \country{Switzerland}
}
\author{Gene Lin}
\affiliation{%
  \institution{Meta Reality Labs}
  \country{Canada}
}
\author{Doug Roble}
\affiliation{%
  \institution{Meta Reality Labs}
  \country{USA}
}
\author{Olga Sorkine-Hornung}
\affiliation{%
  \institution{Meta Reality Labs and ETH Zurich}
  \country{Switzerland}
}
\author{Tuur Stuyck}
\email{tuur@meta.com}
\affiliation{%
  \institution{Meta Reality Labs}
  \country{USA}
}

\renewcommand\shortauthors{Chen, H. et al}

\begin{teaserfigure}
  \centering
  \includegraphics[width=0.95\textwidth]{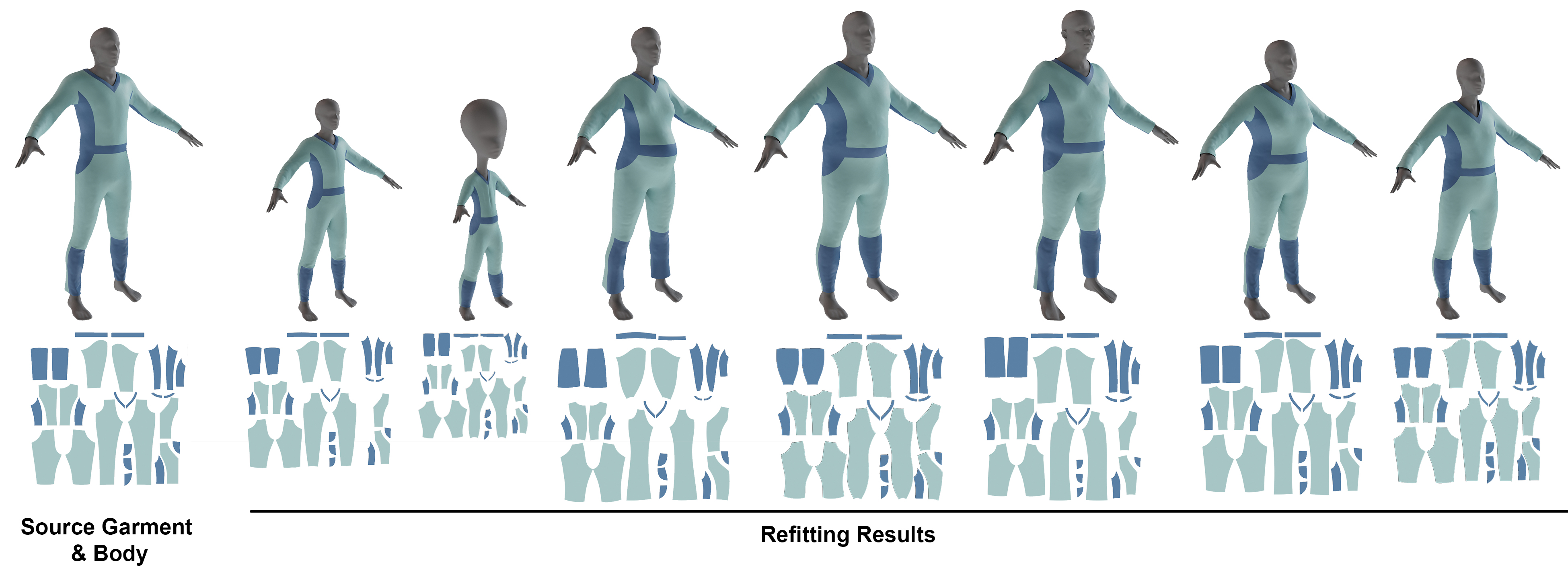}
  \caption{
  We present \emph{Dress Anyone}, a novel, automated method to fit a garment to different body shapes and sizes. We use a differentiable simulator to drape a 3D garment onto a body and adjust the corresponding 2D sewing pattern (bottom row) such that the garment fits appropriately while preserving garment design. As a result, our method produces physically draped garments with their corresponding sewing patterns. Here we demonstrate high quality refitting results for a wide variety of humanoid body shapes for a complex multi-component garment pattern.}
  \label{fig:teaser}
\end{teaserfigure}

\begin{abstract}
Well-fitted clothing is essential for both real and virtual garments to enable self-expression and accurate representation for a large variety of body types. Common practice in the industry is to provide a pre-made selection of distinct garment sizes such as small, medium and large. While these may cater to certain groups of individuals that fall within this distribution, they often exclude large sections of the population. In contrast, individually tailored clothing offers a solution to obtain custom-fit garments that are tailored to each individual. However, manual tailoring is time-consuming and requires specialized knowledge, prohibiting the approach from being applied to produce fitted clothing at scale. To address this challenge, we propose a novel method leveraging differentiable simulation for refitting and draping 3D garments and their corresponding 2D pattern panels onto a new body shape, enabling a workflow where garments only need to be designed once, in a single size, and they can be automatically refitted to support numerous body size and shape variations. Our method enables downstream applications, where our optimized 3D drape can be directly ingested into game engines or other applications. Our 2D sewing patterns allow for accurate physics-based simulations and enables manufacturing clothing for the real world. 
\end{abstract}


%
%


\maketitle

\section{Introduction}

Clothing is an essential aspect of daily life. Given the vast population, the majority of the garment manufacturing industry primarily focuses on producing clothes in a limited series of discrete sizes (e.g. small, medium and large) rather than custom-made pieces for individuals. Although these garments may not provide a perfect fit, they are more economical and satisfactory for a large portion of the population. However, this approach limits the types of fabrics used and is not inclusive of all possible body variations --- certain individuals will require additional tailoring. Similarly, a discrete set of predetermined sizes do not translate well to digital characters where the character body shapes and sizes can vary even more widely. In fact, virtual representations may even be fantastical of nature. Relying on preset clothing sizes will not work on many such characters. While digital artists possess the capability to manually refit these garments, such a task is laborious, time consuming, requiring both access to, and proficiency in specialized software. Therefore, there is a high demand for automated tailoring algorithms that are capable of refitting garments to a wide variety of body types. 

To address this, we propose a computational technique, which creates custom-fitted garments for any humanoid character model by adjusting the 2D sewing patterns of the garment directly. By adjusting the 2D sewing patterns in conjunction with the draped 3D model, rather than solely manipulating the 3D clothing mesh, it becomes possible to replicate these garments physically. It also provides ground-truth rest shape information for downstream applications such as cloth simulation, virtual try-on, fashion design and telepresence applications where garments need to be adapted automatically at scale to various avatar representations.
Therefore, making it a valuable tool for fashion designers and animators alike. Our complete representation of combined physically draped 3D garments and associated 2D patterns also enables other post-processing pipelines such as garment remeshing where important information such as seam line locations can be preserved in the remeshed topology. We demonstrate effectiveness across a diverse range of garments and body shapes ranging from humanoid to fantastical which do not conform to humanoid proportions. We demonstrate the efficacy of our method for both loose and tight-fitting clothing. In summary, our main contributions are
\begin{itemize}[leftmargin=*]
    \item An end-to-end garment refitting method that uses differentiable simulation to produce physically-simulated 3D draped garments, complete with their corresponding 2D sewing patterns. Our approach can refit to different body shapes and sizes with varying body mesh topologies. 
    \item A well-designed control cage formulation for 2D pattern optimization for which we show that it outperforms recent state-of-the-art methods~\cite{wang2018rule, li2023diffavatar}.
    \item A carefully selected combination of loss function components which enable high quality, design preserving results on a variety of clothing items ranging from tight to loose items. 
\end{itemize}
\section{Related Work}

\paragraph{Cloth Simulation} Physics-based simulation of clothing has made incredible strides in the last decades starting with the seminal work of \citet{edm} and that of \citet{BWCloth} on implicit integration enabling stable simulations with large time steps. Since then, many novel methods \cite{fastCloth, choi2002stable} have been proposed such as the optimization formulation of implicit Euler \cite{martin2011example, gast2015optimization}.  Different approaches provide varying trade-offs between stability, speed and accuracy such as Projective~Dynamics~\cite{bouaziz2014projective}, Position Based Dynamics~\cite{muller2007position} and its extension XPBD~\cite{macklin2016xpbd} and most recently Vertex~Block~Descent~\cite{chen2024vertex}. \citet{guo2018material} presented a novel approach leveraging the material point method to simulate thin shells. \citet{stuyck2022cloth} provides an overview.

\paragraph{Differentiable Simulation} There has been a renewed interest in the development of differentiable simulation methods for the purpose of inverse design, system identification~\cite{larionov2022estimating, chen2022virtual} and for integration with learning-based frameworks~\cite{liang2019differentiable} to allow for seamless gradient propagation between simulation models and neural networks. Early work proposed the use of the Adjoint method to differentiate through implicit simulation of clothing~\cite{wojtan2006keyframe} with applications to keyframe control. This idea has been applied to several other simulation frameworks such as DiffXPBD~\cite{stuyck2023diffxpbd} and DiffPD~\cite{du2021diffpd}, which has been extended by DiffCloth~\cite{li2022diffcloth} to address frictional contact for clothing.

\paragraph{Pattern Generation.} High quality clothing patterns are key for accurate clothing simulations as they contain ground truth rest shape information which is often compromised in the 3D drape due to strain in the material. Additionally, they are required for fabrication and they enable custom tailored clothing design using computer systems. Several works have focused on pattern representation and generation using synthetic data. NeuralTailor~\cite{korosteleva2022neuraltailor} presents a learning-based pipeline and a unified model for different garment types, which allows estimating sewing patterns from 3D point clouds. Follow up work GarmentCode~\cite{korosteleva2023garmentcode} presents a programming-based framework for garment pattern construction. DressCode~\cite{he2024dresscode} introduces a GPT-based architecture for generating sewing patterns with text guidance. ISP~\cite{li2024isp} focuses on multi-layered clothing where sewing patterns are represented as signed distance fields.  Given 3D garments, patterns can be extracted using computational methods \cite{pietroni2022computational, bang2021estimating}.

\paragraph{Garment Recovery.} Recently, several advances have been proposed for recovering clothing items from limited real data such as images~\cite{halimi2022pattern, garmentFromImage} and scans~\cite{li2023diffavatar}. Numerous works~\cite{chen2024panelformer,liu2023towards,chen2022structure} present deep neural networks to parameterize the space of garment sewing patterns, which allows them to predict sewing patterns from images of clothing. Similarly, \citet{li2023garment} introduce a fitting method that leverages shape and deformation priors derived from synthetic data to obtain 3D garment reconstruction from static images. \citet{sarafianos2024garment3dgen} presents a method to generate simulation-ready 3D clothing from images or text prompts leveraging generative neural networks. DiffAvatar~\cite{li2023diffavatar} uses differentiable simulation to optimize the 2D garment pattern to recover clothing assets from static scans of clothed people. Follow up work PhysAvatar~\cite{PhysAavatar24} extends this work to enable avatar recovery from multi-video data. A separate line of work focuses on garment appearance recovery \cite{xiang2022dressing}.

\paragraph{Garment Refitting.} A popular line of research focuses on refitting garments from one body shape to another target shape. Some methods operate on the 3D shape of the garment directly~\cite{fernando} leveraging an iterative optimization approach. \citet{brouet2012design} proposed a geometric constrained optimization problem that produces refitted 2D patterns but does so without considering simulated draping effect. \citet{wang2018rule} also uses differentiable simulation to optimize for 2D patterns, but its functionality has only been demonstrated to work on a smaller range of body shape variations compared to our work. \citet{bartle2016physics} proposed a 2D pattern optimization procedure which enables direct edits to the 3D garment geometry. Recent work considers body movement and its effect on personalized garment fits \cite{wolff2023designing}.

Compared to prior research, our method is the first to demonstrate refitting with corresponding 2D patterns, taking into account physical drape across a broad spectrum of body shapes, rather than a limited set of pre-determined template bodies.



\section{Background}

We use differentiable cloth simulation to drape a 3D garment on a static body shape in A-pose. This allows us to obtain a physically-based drape that is in an equilibrium state resting on the body under external forces. The rest shape of the 3D triangles are provided by the 2D sewing pattern. We provide a concise overview of garment simulation, including its differentiable formulation and the subsequent application in optimizing 2D sewing patterns.

\subsection{Garment Simulation}


Our contributions can be used in combination with any differentiable simulation method. Here, we use XPBD~\cite{macklin2016xpbd} to produce the garment drapes on a given body shape because it provides fast and stable results. 

We formulate the energy $U(\mathbf{x})$ as a set of constraints functions $\mathbf{C} = \left[C_1(\mathbf{x}), \cdot \cdot \cdot, C_m(\mathbf{x})\right]^{\top}$ and inverse compliance matrix \(\bm{\alpha}^{-1}\) as
\begin{equation}
U(\mathbf{x}) = \frac{1}{2} \mathbf{C}(\mathbf{x})^{\top} \bm{\alpha}^{-1} \mathbf{C}(\mathbf{x}).    
    \label{eq:xpbdUenergy}
\end{equation}
The system is solved iteratively with the introduction of constraint multipliers $\bm{\lambda}$ which is computed under the implicit Euler time scheme as
\begin{align}
    (\nabla \mathbf{C}(\mathbf{x})^{\top}\mathbf{M}^{-1}\nabla\mathbf{C}(\mathbf{x}) + \tilde{\bm{\alpha}}) \Delta \bm{\lambda} = - \mathbf{C}(\mathbf{\mathbf{x}}) - \tilde{\bm{\alpha}}\bm{\lambda}, \label{eq:xpbd}
\end{align}
where $\tilde{\bm{\alpha}} = \bm{\alpha} / \Delta t^2$, and $\mathbf{M}$ is the mass matrix. Given $\Delta \bm{\lambda}$, the position update is then computed as
\begin{align}
\Delta \mathbf{x} = \mathbf{M}^{-1} \nabla \mathbf{C}(\mathbf{x})\Delta \bm{\lambda}
\end{align}
With external forces denoted by $\mathbf{f}_{\text{ext}}$ acting on the system, the state $\mathbf{q}_n = \left(\mathbf{x}_n,\mathbf{v}_n\right)$ at time step $n$ consisting of positions $\mathbf{x} \in \mathbb{R}^{3V}$ and velocities $\mathbf{v} \in \mathbb{R}^{3V}$ is updated as
\begin{equation}
\begin{aligned}
 \mathbf{x}_{n+1} 
 &= \mathbf{x}_{n} + \Delta \mathbf{x}\left(\mathbf{x}_{n+1} \right) + \Delta t \left( \mathbf{v}_n + \Delta t \mathbf{M}^{-1} \mathbf{f}_{\text{ext}} \right) \\
 \mathbf{v}_{n+1} 
 &= \frac{\tau}{\Delta t} \left( \mathbf{x}_{n+1} - \mathbf{x}_{n} \right)
\end{aligned}
\label{eq:xpbdUpdateRule}
\end{equation}
Note that compared to the original formulation, we incorporate additional velocity damping to ensure that the garment attains a stable drape on the body, where $\tau$ is the velocity damping coefficient, which is set to 0.95 in our simulations.

\subsection{Differentiable Simulation}
Our work largely follows the structure proposed in DiffXPBD~\cite{stuyck2023diffxpbd} which provides a differentiable formulation of the XPBD simulation method. To optimize the loss function $\mathcal{L}$ over the control variables $\mathbf{u}$, we compute the gradient through the full dynamic sequence. We leverage the Adjoint method for efficient gradient computation by first computing intermediate Adjoint states $\hat{\mathbf{q}}_n= (\hat{\mathbf{x}}_n \in \mathbb{R}^{3V}, \hat{\mathbf{v}}_n \in \mathbb{R}^{3V} )$ for all simulation steps N. Let $\mathbf{Q} = \left[\mathbf{q}_0, \cdot \cdot \cdot, \mathbf{q}_m\right]$ be the concatenation of states $\mathbf{q}_n =  (\mathbf{x}_n, \mathbf{v}_n)$ over time. Then, we can represent the advancement of the state as $\mathbf{Q} = \mathbf{F}(\mathbf{Q},\mathbf{u})$, where $\mathbf{F}$ contains all of the time step formulae from \myeqref{eq:xpbdUpdateRule}.

Following the Adjoint method, the gradient is computed as
\begin{equation}
    \frac{d \mathcal{L}}{d \mathbf{u}} = \hat{\mathbf{Q}}^{\top} \frac{\partial \mathbf{F}}{\partial \mathbf{u}} + \frac{\partial \mathcal{L}}{\partial \mathbf{u}},
    \label{eq:gradientRule}
\end{equation}
where $\hat{\mathbf{Q}}$ is the concatenation of all Adjoint states $\hat{\mathbf{q}}_n$. Applying \myeqref{eq:gradientRule} to our modified XPBD integration scheme  in \myeqref{eq:xpbdUpdateRule} results in a slight variation of the Adjoint state computation 
\begin{equation}
\begin{aligned}
    \hat{\mathbf{x}}_{n} 
    &= \hat{\mathbf{x}}_{n+1} + \left(\frac{\partial \Delta \mathbf{x}}{\partial \mathbf{x}}\right)^\top \hat{\mathbf{x}}_{n}
    + \frac{\tau \hat{\mathbf{v}}_n}{\Delta t}
    - \frac{\tau \hat{\mathbf{v}}_{n+1}}{\Delta t}
    + \frac{\partial \mathcal{L}}{\partial \mathbf{x}}^\top \\
    \hat{\mathbf{v}}_n 
    &= \Delta t \hat{\mathbf{x}}_{n+1} + \frac{\partial \mathcal{L}}{\partial \mathbf{v}}^\top,
\end{aligned}
\label{eq:rawAdjoint}
\end{equation}
where we assume the external forces are not dependent on the position and there are no velocity dependent energy terms. Once all Adjoint states have been computed using \myeqref{eq:rawAdjoint}, we can compute the gradient with respect to the control variable \(\mathbf{u}\) as presented in \myeqref{eq:gradientRule}. In practice, this entails computing and storing $\frac{\partial \Delta \mathbf{x}}{\partial \mathbf{u}}$ for any control variable for which gradients are required.


\section{Method}


\begin{figure*}
    \centering
     \includegraphics[width=0.95\textwidth]{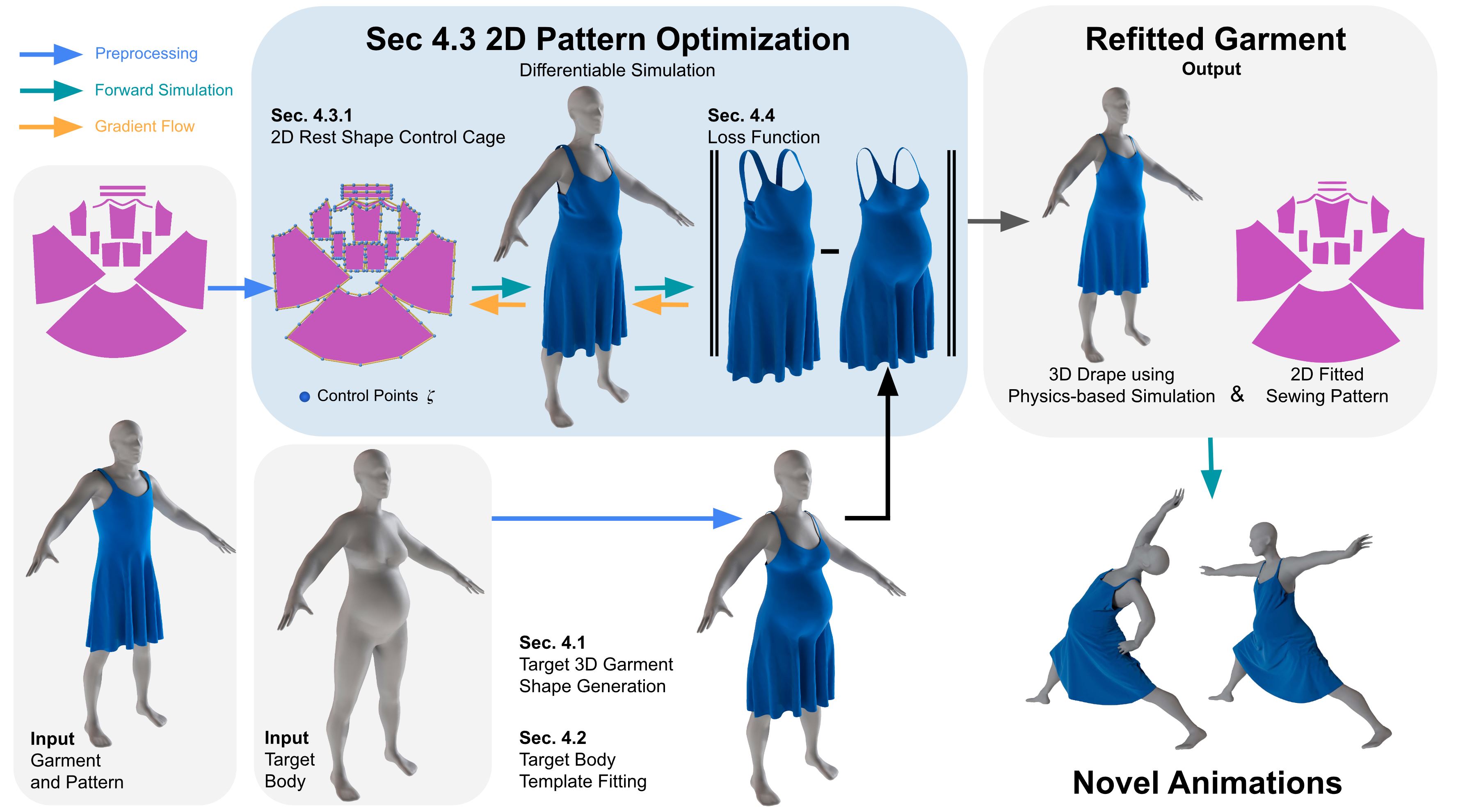}
     \caption{Given a draped input garment and body, \emph{Dress Anyone} produces refitted 2D patterns and a 3D draped garment to fit a provided target body shape. We first fit a template body model to the target body to make the method robust against topology differences between body meshes. We the produce a target 3D garment shape \cite{fernando} which we use as a target shape. We then use differentiable simulation to optimize for a refitted 2D pattern that fits the target body. We leverage a robust and efficient control cage formulation which preserves the garment design. Our refitted garments can be used for several downstream applications such as novel motion generation.
     }
     \label{fig:pipeline}
\end{figure*}

Given a reference garment designed and fitted to a specific body shape and its corresponding 2D sewing pattern, our method automatically optimizes for a custom fitted and physically simulated 3D garment, draped on a new body shape along with its corresponding and optimized 2D sewing patterns. We first generate a target 3D shape for the garment item leveraging prior work~\cite{fernando} (Sec.~\ref{sec:autofit}) to serve as an optimization target later in the method. We propose an additional statistical body model fitting step to alleviate limitations in prior work (see Sec.~\ref{sec:bodyFitting}). Given this target shape, we employ differentiable simulation to optimize for the 2D sewing pattern such that the 3D garment, which is draped using differentiable physics-based simulation, matches the target shape as closely as possible (see Sec.~\ref{sec:optimization}). The 2D sewing pattern is parameterized using a number of control points on the sewing patterns that enforce conformal changes with every update. Our choice of pattern representation significantly aids in regularizing the deformation space to preserve the shape and design intent of the sewing patterns of the garment when being custom fit to novel body shapes and sizes. Fig.~\ref{fig:pipeline} visualizes the different components of our computational method.

\subsection{Target Body Template Fitting}
\label{sec:bodyFitting}
The target shape generation method~\cite{fernando} requires a common mesh connectivity between body geometries. However, meshes can come from a variety of asset generation sources with varying topologies. Therefore, we fit a statistical body model~\cite{SMPL} to the target body shape, making our proposed method compatible with novel body topologies. Given the recovered model parameters, we pose the fitted body to the canonical A-pose.

\subsection{Target Shape Generation}
\label{sec:autofit}
Provided with a garment fitted to a certain body, we first estimate a 3D target shape of the refitted garment to a novel body without accompanying 2D sewing patterns using the refitting pipeline from~\citet{fernando}.  Note that this method only provides a 3D shape without physics-based simulation, which means the garment is not
in any draped equilibrium state;
even more importantly, it does not produce the required 2D patterns. This 3D target shape is subsequently used in our differentiable pipeline as the target drape for which we optimize the garment sewing patterns. Note that because the target shape is not physically simulated, we do not expect to match this shape exactly.

\subsection{Sewing Pattern Optimization}
\label{sec:optimization}
Our method optimizes for the 2D sewing pattern by minimizing a loss in 3D space through the use of differentiable simulation. For our initial estimate, we establish a global scaling factor by computing the average scale between the area of the target shape and the reference pattern area across each triangle. We then uniformly scale the reference pattern using this factor. At every iteration, we drape the current best estimate of the sewing pattern onto the body and simulate until equilibrium is reached. This allows us to evaluate our loss formulation (Sec.~\ref{sec:loss}), which compares the simulated result to our target shape (Sec.~\ref{sec:autofit}). After loss evaluation, we can compute the gradient of the final simulated state of the 3D vertices with respect to the 2D sewing pattern vertices. 
This quantity can then be transformed to the control cage vertices, which allows us to update our best estimate through the use of gradient descent.

\subsubsection*{Rest Shape Control Cage} 
Although it would be possible to optimize for the 2D sewing pattern vertices directly, this would produce noisy results which do not preserve the original design of the garment as documented by \citet{li2023diffavatar}. Instead, it is desired to regularize the optimization representation through the use of a lower-dimensional control cage as illustrated in Fig.~\ref{fig:pipeline}. Observing that the boundary of the sewing patterns are composed of smooth curves and to constrain the parametric space, we propose to use Green coordinate~\cite{lipman2008green} control points to modify the change in the pattern space. Our selection of the Green coordinate control cage, in contrast to the Harmonic control cage~\cite{wang2018rule} and the mean value coordinate control cage~\cite{li2023diffavatar}, imposes restrictions on the degrees of freedom, requiring the patterns to deform in a smooth and conformal way. The pattern is enclosed by the control points such that there are no discontinuity at the boundaries. Given $C$ control points $\bm{\zeta} \in \mathbb{R}^{3C}$ and $V$ vertices, where $C \ll V$, we can express the vertex positions  $\Bar{\mathbf{x}}$ on the 2D sewing patterns as $\Bar{\mathbf{x}} = \bm{W}_1 \bm{\zeta} + \bm{W}_2 \mathbf{n}(\bm{\zeta})$, 
where $\mathbf{n}$ is the outward normal of the edge of the control cage, and $\mathbf{W}_1$ and $\mathbf{W}_2$ are $3V \times 3C$ constant weight matrices. We compute the gradient as in \myeqref{eq:gradientRule}. Therefore, we need to compute \(\frac{\partial \mathbf{F}}{\partial \mathbf{\zeta}}\), which requires us to compute \(\frac{\partial \Delta \mathbf{x}}{\partial \bm{\zeta}}\). The gradient of each control point is the sum of the weighted gradient from the interior points. We thus find 
\begin{equation}
    \frac{\partial \Delta \mathbf{x}}{\partial \bm{\zeta}} =\frac{\partial \Delta \mathbf{x}}{\partial \Bar{\mathbf{x}}}\left(  \bm{W}_1 + \bm{W}_2 \frac{\partial \mathbf{n}}{\partial \bm{\zeta}}\right)
\end{equation}
where $\frac{\partial \Delta \mathbf{x}}{\partial \Bar{\mathbf{x}}}$ is computed by differentiating through the position update of the in-plane stretching and shearing formulation of the cloth triangle constraint. 

\subsection{Loss}
\label{sec:loss}
We propose a weighted combination of several loss components,
each with its own purpose. We leverage feature matching terms to match the provided 3D target shape as closely as possible. However, doing so without any additional regularizing terms produces low quality garment patterns that no longer preserve the original design of the garment. To produce desired results, we augment our loss function with several regularizing terms.

\subsubsection{Target Shape Matching Terms}
We observe that the position of the boundary and seam lines play a crucial role in determining the style and fit of an outfit, such as the shoulder line. These elements serve as vital indicators for assessing the quality of a refitted outfit. Consequently, our method strives to minimize the discrepancy in the 3D positions of the boundary and seam lines between the target shape and the simulated result. 
In addition, to maintain the fit, our loss functions impose penalties on the difference between the interior points. We aim for a close match at the boundary and the seams. However, given that the target shape is not physically simulated, we allow for some slack in the interior by applying a smaller weight. 
The loss is formulated as
\begin{equation}
    \mathcal{L}_{SM} = \alpha \sum_{x \in \partial \Omega}\lVert \mathbf{x} - \mathbf{x}^* \rVert^2 + \beta \sum_{x \in \text{Seam}}\lVert \mathbf{x} - \mathbf{x}^* \rVert^2 + \gamma \sum_{x \in \Omega}\lVert \mathbf{x} - \mathbf{x}^* \rVert^2,
\end{equation}
where $\mathbf{x}^*$ is the target position, and $\alpha$, $\beta$, and $\gamma$ are the weights, with $\gamma \ll \alpha, \beta$.

\subsubsection{Boundary Curvature Term}
Although the Green coordinates guarantee conformal changes of the interior, the curvature at the boundary can be distorted under large displacement of the control points and deviates from the reference design. To alleviate the problem, we impose an additional loss to penalize the change of the curvature on the boundary vertices~\cite{li2023diffavatar, wang2018rule}. We seek a scaled rotation matrix $\mathbf{T}_i = s \mathbf{R}_i \in \mathbb{R}^{2 \times 2 }$ at each point $\mathbf{x}_i$ with least curvature distortion to its connected boundary edges, $\mathbf{T}_i = \arg \min_{\mathbf{T}} || \mathbf{e}_{i1} - \mathbf{T} \Bar{\mathbf{e}}_{i1} ||^2 + || \mathbf{e}_{i2} - \mathbf{T} \Bar{\mathbf{e}}_{i2} ||^2$ with $\mathbf{e}_{i1} = \mathbf{x}_{i + 1} - \mathbf{x}_i$ and  $\mathbf{e}_{i2} = \mathbf{x}_{i - 1} - \mathbf{x}_i$. The loss is defined as the accumulation of the curvature distortion as 
\begin{equation}
\mathcal{L}_\text{curvature} = \sum_{i \in \partial\Omega} \| (\mathbf{e}_{i1} - \mathbf{T}_i \Bar{\mathbf{e}}_{i1}) + (\mathbf{e}_{i2} - \mathbf{T}_i \Bar{\mathbf{e}}_{i2}) \|^2,
\end{equation}

\subsubsection{Pattern Matching Term}


The optimization on each panel are done independently, and can lead to mismatch pattern boundaries. To ensure the pieces can be sewn together without artifacts such as cloth gathering at the seams, we introduce an additional loss term to enforce the boundary of the corresponding sewing pieces to have the same length.
\begin{equation}
\mathcal{L}_\text{PM} = \sum_{i \in \text{Seam edges}} \| \mathbf{x}_{i} - \mathbf{x}_{i+1} ||^2 - || \mathbf{x}'_i - \mathbf{x}'_{i+1}\|^2
\end{equation}

\subsubsection{Total Area Loss}
We include an additional area term in our loss, which measures the difference of the target total surface area and the total surface area of the simulated shape of each panel. This term is especially important for loose fitting garments where the gradient of the interior points becomes noisy due to wrinkles and folds. 
\begin{equation}
    \mathcal{L}_{TA} = \sum_{p \in \text{panels}}(\text{A}_{p} - \sum_{i \in \text{T}_p} \text{A}_i)^2
\end{equation}

\subsection{Pattern Symmetry}

Most garments are designed to be symmetric, with flip symmetry in the sewing patterns. When desired, we enforce symmetry during our optimization process. We first detect the corresponding pairs of patterns with flip symmetry, and enforce the update to be the average of the gradient on the pair of control points. However, most people have some degree of body asymmetry and therefore, truly custom fits may require intentional asymmetry in garment design. 

\section{Results}

\begin{figure*}
    \centering
     \includegraphics[width=0.99\textwidth]{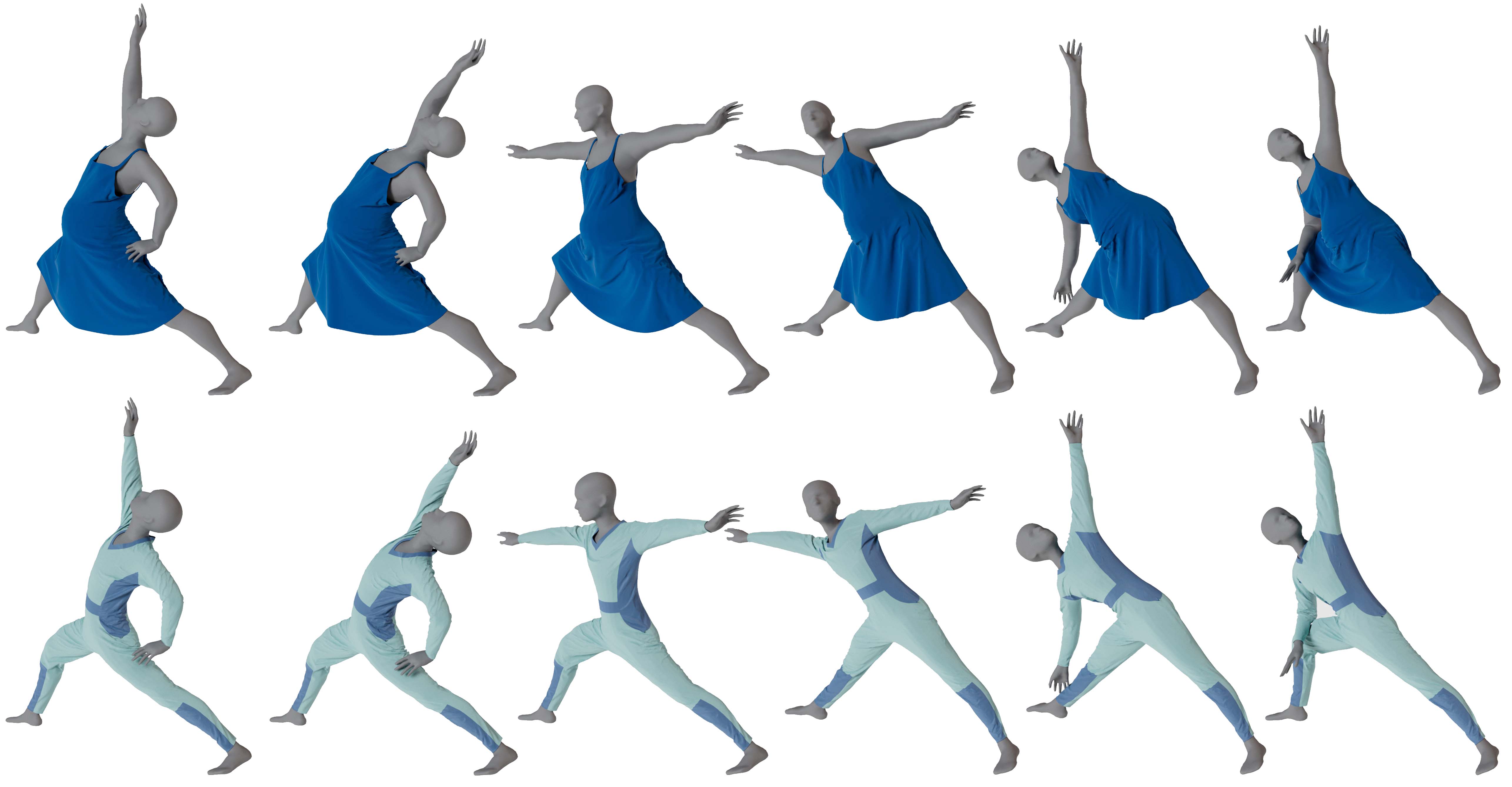}
     \caption{\emph{Animated sequences of refitted clothing.} Our method produces custom fit garments, allowing them to be directly draped and simulated. The incorporation of 2D sewing patterns provides an accurate rest shape which significantly improves garment simulation realism. 3D draped geometries do not provide an accurate rest shape because they frequently contain deformations such as stretching or sagging under gravity which are already integrated into the mesh. This can lead to inaccuracies during the simulation process. Additionally, it allows for our garments to be manufactured from real fabrics. We showcase a select number of frames of a yoga sequence, demonstrating that the fitted garments allow for rich dynamics and diversity in body pose.}
     \label{fig:NewMotion}
\end{figure*}

We evaluate our method by refitting various clothing items to a wide variety of body shapes. All results are obtained with symmetry enforcement. 
Our results demonstrate that our method is capable of generalizing to new body shapes, where refitted garments retain their original design intend. We produce simulation-ready garment assets, which enables downstream applications such as the novel animations shown in Fig.~\ref{fig:NewMotion}.

\begin{figure}
\centering
    \includegraphics[width=0.5\textwidth]{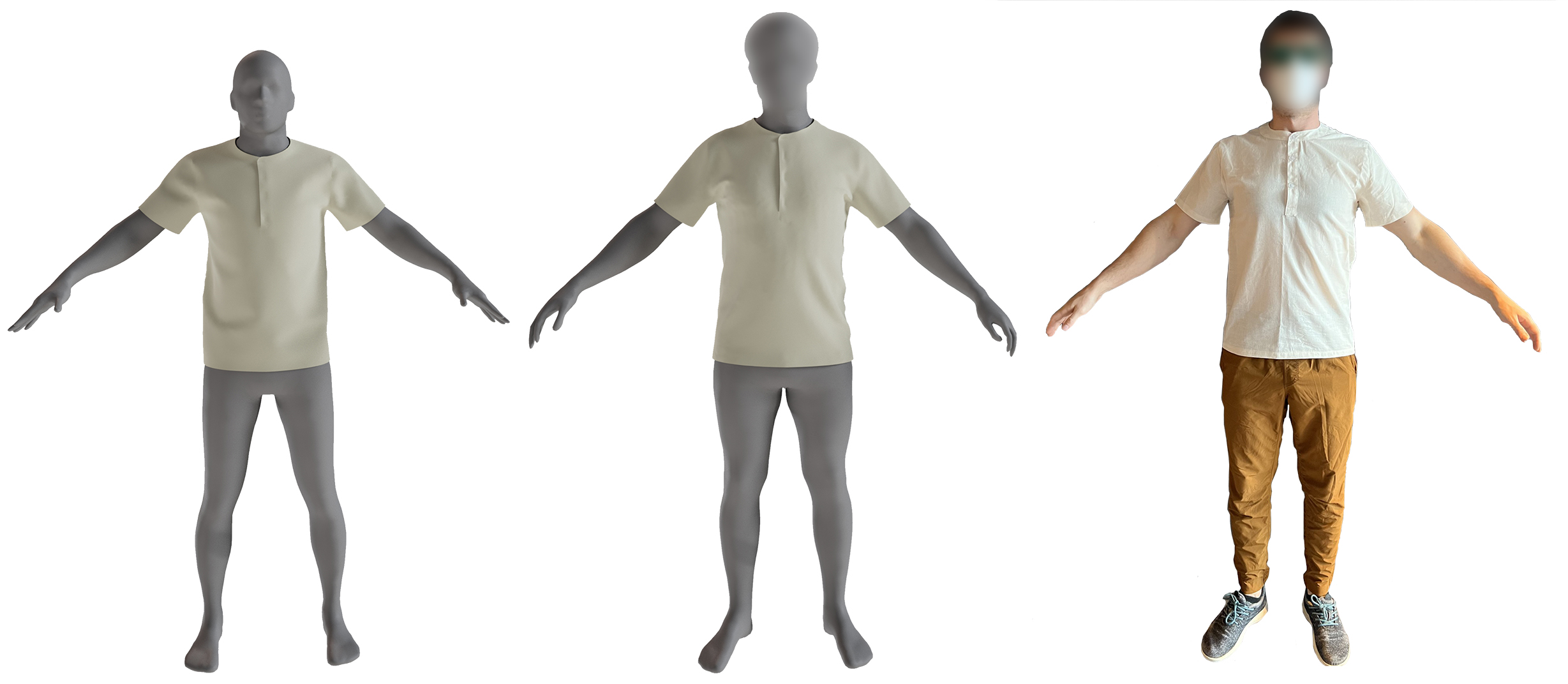} 
    \caption{We demonstrate a real world example of a garment refitted using our approach where we manufactured a physical shirt given the refitted sewing patterns. Left shows the original garment fitted to the original body. Middle shows our virtual refitted garment which is physically draped on the new body. The body geometry was obtained using a 3D scan. Right shows the real garment draped on the real person from whom the scan was taken. Note how the real drape matches the virtual resized garment.
    }
    \label{fig:realShirt}
\end{figure}

\begin{figure*}
    \centering
     \includegraphics[width=0.85\textwidth]{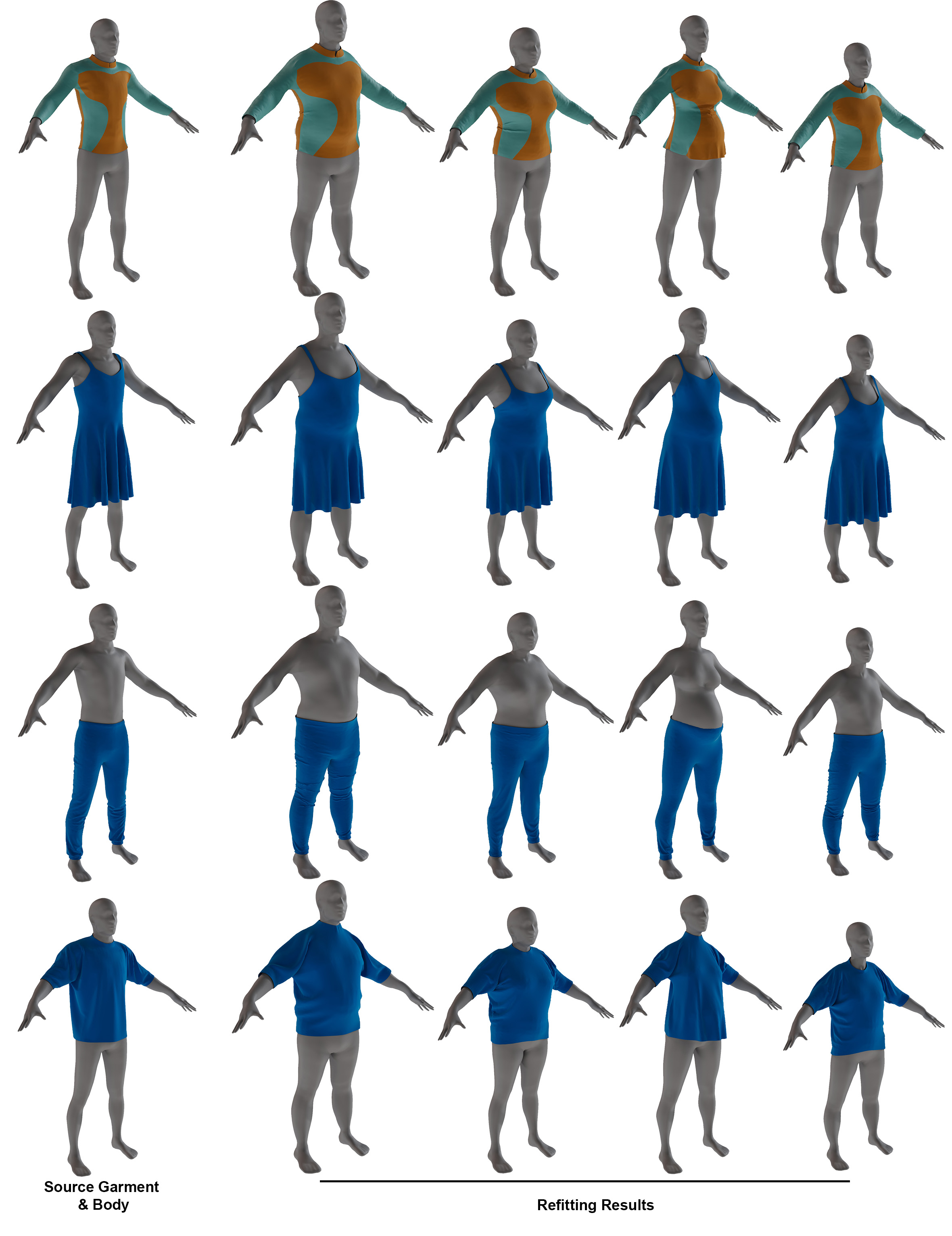}
    \caption{\emph{Optimization Results.} We showcase a collection of refitted garments onto different body shapes. We demonstrate that our automatic computational method is effective at fitting garments for numerous garment types, both loose and tight fitting, without requiring any garment specific modifications or tuning. We demonstrate results on a wide range of body types and sizes.}
    \label{fig:garmentCatalog}
\end{figure*}

Figures~\ref{fig:teaser}~and~\ref{fig:garmentCatalog} demonstrate results for a variety of garments, both tight and loose fitting, refitted onto different body shapes and sizes. Note that our method is capable of handling complex garment designs. The garment in Fig.~\ref{fig:teaser} consists of a total of 28 individual sewing pattern panels. We show examples of refitting garments with significantly non-uniform changes in the body shapes, see for example the pregnant woman in Fig.~\ref{fig:garmentCatalog} or the alien in Fig.~\ref{fig:teaser} with disproportionate limb sizes where the arms are much shorter compared to the reference. Additionally, notice that in Fig.~\ref{fig:comparison}, the curvy body has a much smaller waist line compared to their upper torso. Yet, our method produces a properly refitted pattern to generate a custom fit. 
To showcase that our approach produces manufacturable clothing items, we created a T-shirt garment from refitted patterns provided by our method and compare the virtual drape to the real drape in Fig.~\ref{fig:realShirt}. The shirt is made from non-stretchable fabric and would highlight any areas of poor fit. However, our result shows that it aligns very well with the target body.

\begin{table}[t]
    \centering
    \setlength{\tabcolsep}{0.5mm}
    \caption{\textbf{Quantitative Comparisons}. 
    Measuring a triangle quality indicator~\cite{shewchuk2002quality}, we find that we produce the best results for the refitted sewing patterns shown in Fig.~\ref{fig:comparison}. Higher values indicate better quality.}
    \vspace{-0.2cm}
    \resizebox{0.6\columnwidth}{!}{
    \begin{tabular}{lp{1.7cm}p{1.7cm}p{1.7cm}c}
        \toprule
         & \centering Ours & \centering DiffAvatar & \centering \cite{wang2018rule} & \\
        \midrule
        Min $\uparrow$ & \centering \textbf{0.263922} & \centering 0.195618 & \centering 0.10746 & \\
        Avg $\uparrow$ & \centering \textbf{0.944209} & \centering0.938846 & \centering 0.944205 & \\
        \bottomrule
    \end{tabular}
    }
    \label{tab:CDMeasurements}
\end{table}

\subsection{Comparison to Related Work}

\begin{figure*}
    \includegraphics[width=0.95\textwidth]{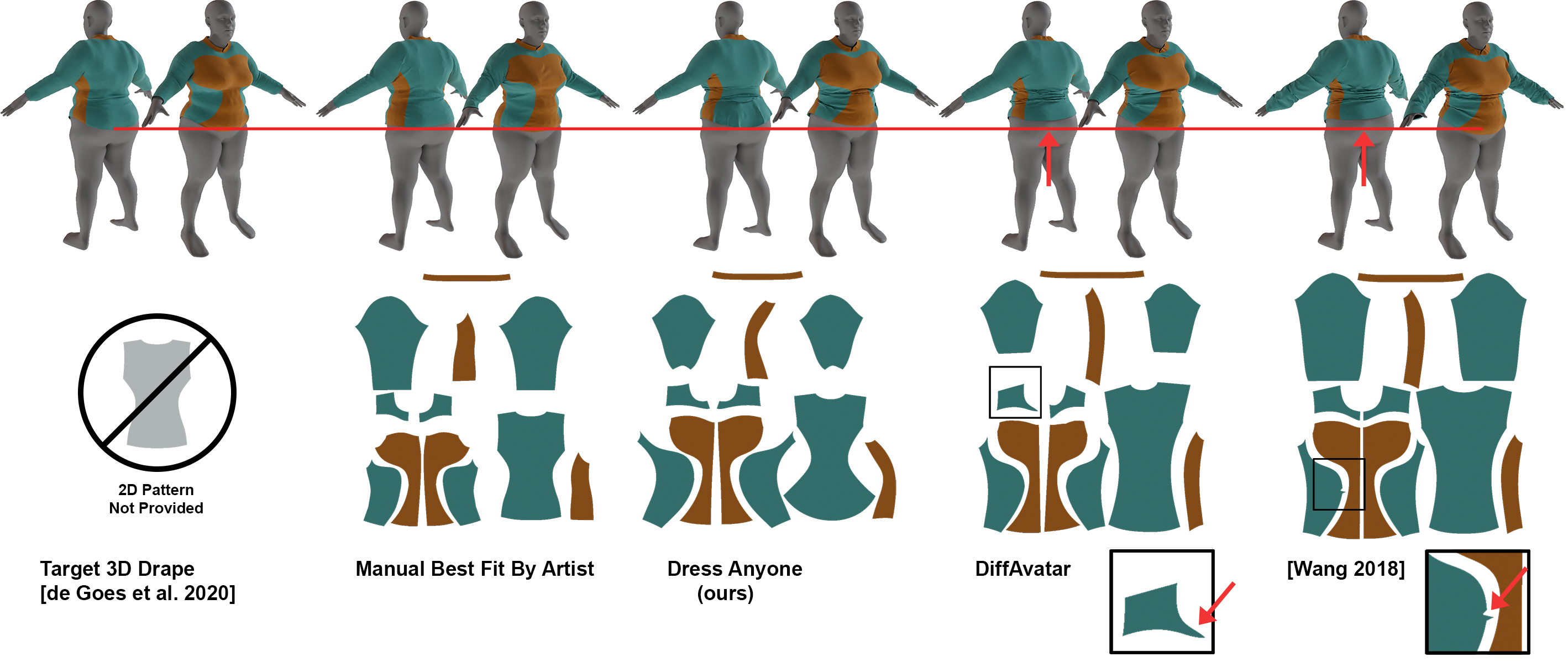} 
    \caption{\emph{Comparison to Related Work.} We compare our proposed method with several other recent state-of-the-art techniques. For reference, we also showcase a manually obtained result, produced by a skilled artist using professional specialized tools. Note that out of all computational methods, our method most closely matches the desired 3D target shape shown on the far left. The mismatch is highlighted by the red line and arrows in the top row. The enlarged inset figures highlight that DiffAvatar produces slim elongated features and \citet{wang2018rule} produces noisy features. Both produce sub optimal triangle meshes as highlighted by a quantitative comparison on triangle mesh quality provided in Table~\ref{tab:CDMeasurements}.
    } 
    \label{fig:comparison}
\end{figure*}

Our approach is most similar to DiffAvatar but differs in several key aspects. DiffAvatar is designed to fit template garments to incomplete scans of clothed people whereas ours provides an end-to-end method to refit garments onto any body shape producing a 3D drape with associated 2D sewing patterns. Our approach provides better regularization through our choice of control cage formulation and enables symmetry in the produced patterns, resulting in more realistic sewing patterns. As a result, our method's capability to refit a wider array of body shapes is improved, as demonstrated in Fig.~\ref{fig:comparison}. Using the same loss function, we optimized the sewing patterns with the use of the three different control cages, and only our proposed Green coordinate control cage finds a desirable high-quality result. Since the other two control cages do not guarantee conformal changes, the updated patterns often result in shapes that deviate from the reference design and introduce poorly-shaped narrow triangles, which can lead to numerical instability for both the forward and backward simulation, see Tab.~\ref{tab:CDMeasurements} for a quantitative comparison. As a result these approaches cannot further reduce the loss. We provide a comparison of the 2D pattern quality using the triangle quality metric provided by~\citet{shewchuk2002quality}, and the result shows that our method has the best quality, which is important to physically simulate the refitted garment. 

\subsection{Ablation Studies}

\begin{figure}
    \centering
    \includegraphics[width=0.5\textwidth]{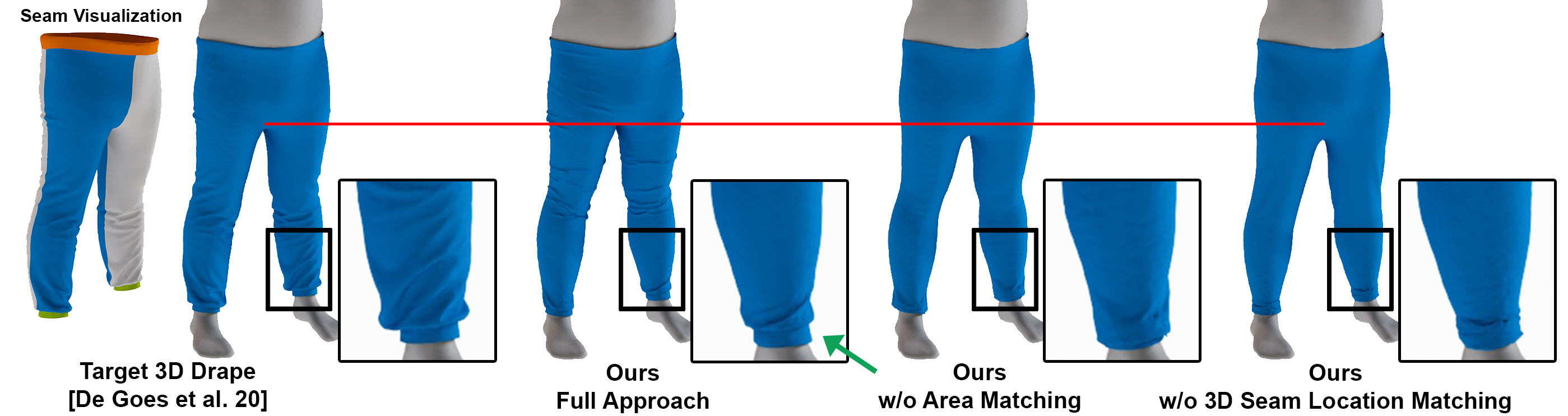} 
    \caption{\emph{3D Ablation.} We provide a visual ablation of the resulting 3D physically simulated drapes when omitting loss term regularizers. Far left shows a visualization of the individual garment panels, highlighting the seam locations used for the seam location matching. We then show the target 3D shape obtained by \citet{fernando} which does not produce 2D sewing patterns. Our full method closely matches the target drape, including the faithful recreation of the ankle cuffs, with custom fit sewing patterns. Note that we do not expect an exact match since the target drape is not physically simulated under external forces such as gravity or body and cloth self collisions. Omitting either of the regularizers produces lower quality fits that do not match the target as closely. Note especially that the ankle cuffs are only preserved with our full approach. The target location of the crotch seam line (highlighted by the red line) is only matched with our full method.
    }
    \label{fig:3Dablation}
\end{figure}
\begin{figure}
\centering
    \includegraphics[width=0.49\textwidth]{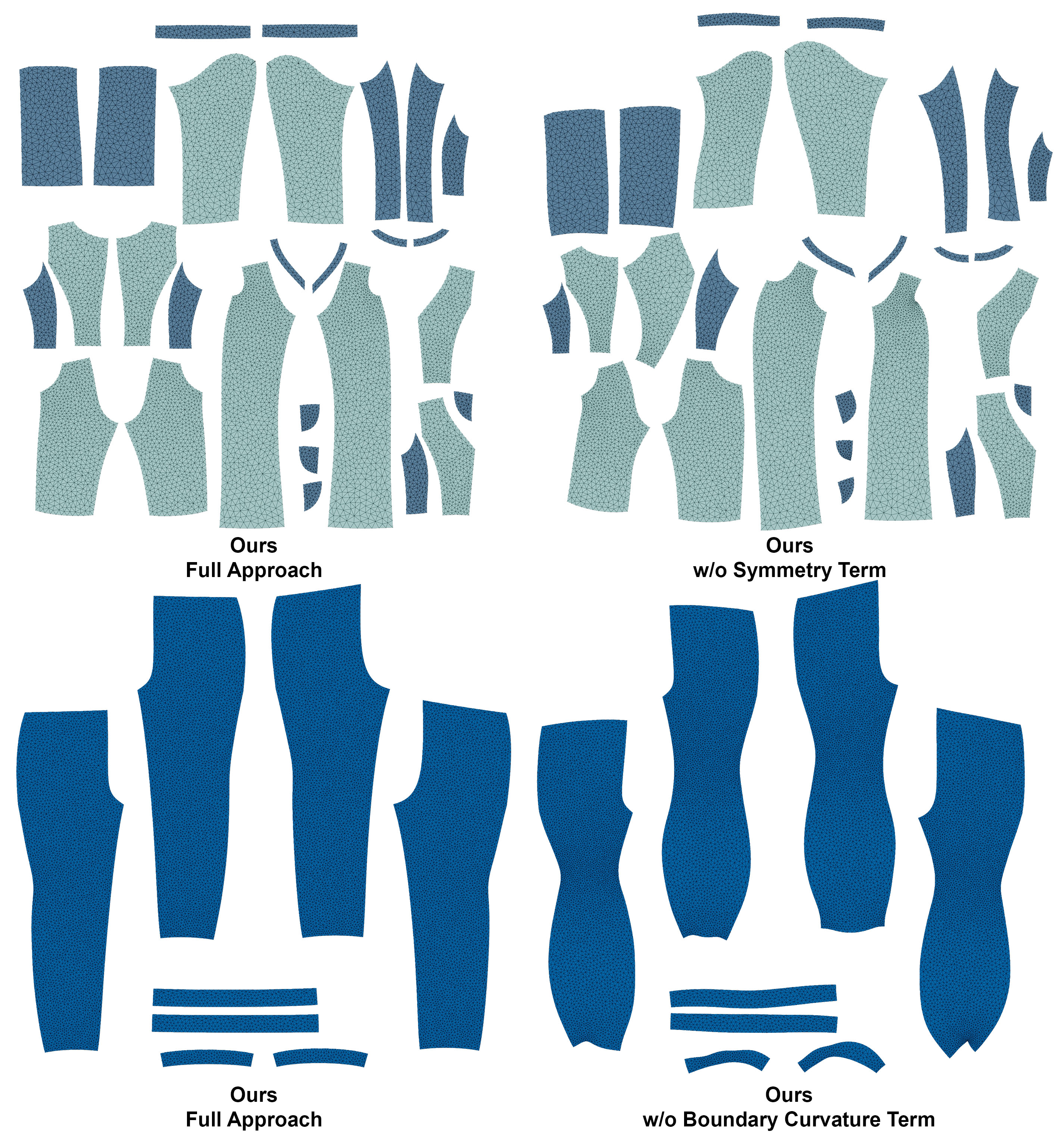} 
    \caption{\emph{2D Ablation.} We assess the importance of our proposed symmetry enforcing approach (ski suit, top) and boundary curvature regularization (pants, bottom) by analyzing the sewing patterns in 2D space. Top : Symmetry enforcement naturally leads to realistic garments patterns to how an experienced tailor would produce them. Bottom : We additionally see the importance of boundary curvature regularization. Omitting this term produces curved pattern boundaries, drastically changing the intended design of the original fit.
    }
    \label{fig:2Dablation}
\end{figure}

We validate our method with ablation studies to verify that the terms in Sec.~\ref{sec:loss} assist in producing high quality refits. We demonstrate an ablation of the seam line vertex matching term and the total area matching term on the 3D results in Fig.~\ref{fig:3Dablation}. This shows that without these terms, the optimization converges to a suboptimal state where it cannot reach the reference shape at the bottom of the pants.
At the top of Fig.~\ref{fig:2Dablation}
we compare the 2D patterns with and without enforcing symmetry on the patterns, and it is evident that the patterns are more desirable with the enforced symmetry. Lastly, at the bottom of Fig.~\ref{fig:2Dablation}, we compare the patterns with and without the loss term on the boundary curvature. Note that while Green coordinates guarantee conformal mapping and thus maintain good triangle quality at largely deformed areas, without the boundary curvature loss the pattern contains less pleasant geometry with a largely curved boundary that is uncommon for regular patterns. 

\subsection{Performance and Implementation Details}

We implemented our algorithm in C++. All experiments were run on CPU using the AMD Ryzen Threadripper PRO processor. The method takes between 1 to 5 minutes for each optimization iteration, which includes forward simulation and gradient back propagation. The total refitting time depends on optimization iteration count and is typically between 5 to 30 minutes depending on the mesh resolution of the garment.
\section{Limitations and Future Work}

The scope of this work is limited to single layer garments. However, we believe our method to be directly applicable to multi-layer garments provided that the differentiable simulator is able to accurately resolve all cloth interactions. Due to the refitting approach of our method, we preserve the garment design, a limitation of this method is that we can not adapt to virtual characters where additional pattern pieces are required such as adding additional sleeves.

As future work, we would like to include differentiable soft-tissue simulation of the underlying flesh to additionally optimize for comfort when wearing the garments. A natural extension would be to include several different body poses in the optimization loss in order to obtain a better fit under movements as proposed by~\citet{wolff2023designing}. Our proposed end-to-end differentiable physics-based method allows for the direct integration of such physics-based losses into the computational framework. 

\section{Conclusion}

We introduced Dress Anyone, a novel computational approach that leverages differentiable simulation and a well-suited pattern representation for optimization to improve on the long standing problem of automated tailoring. Our method incorporates 3D physics-based draping and rest shape optimization to obtain a full representation of garment assets enabling automatic processing for downstream applications. Garments need to be designed only once and can be reused for different body shapes and sizes, including fantastical body types. Our method allows artists to focus their creativity on garment design without burdening them with the laborious and repetitive task of garment refitting.

\section{Acknowledgements}
We would like to thank Romain Pr\'evost for providing us with an implementation of \citet{fernando} and Natasha Devaud for their help with garment design and manufacturing.

\bibliographystyle{ACM-Reference-Format}
\bibliography{ref}


\begin{thebibliography}{43}


\ifx \showCODEN    \undefined \def \showCODEN     #1{\unskip}     \fi
\ifx \showDOI      \undefined \def \showDOI       #1{#1}\fi
\ifx \showISBNx    \undefined \def \showISBNx     #1{\unskip}     \fi
\ifx \showISBNxiii \undefined \def \showISBNxiii  #1{\unskip}     \fi
\ifx \showISSN     \undefined \def \showISSN      #1{\unskip}     \fi
\ifx \showLCCN     \undefined \def \showLCCN      #1{\unskip}     \fi
\ifx \shownote     \undefined \def \shownote      #1{#1}          \fi
\ifx \showarticletitle \undefined \def \showarticletitle #1{#1}   \fi
\ifx \showURL      \undefined \def \showURL       {\relax}        \fi
\providecommand\bibfield[2]{#2}
\providecommand\bibinfo[2]{#2}
\providecommand\natexlab[1]{#1}
\providecommand\showeprint[2][]{arXiv:#2}

\bibitem[Bang et~al\mbox{.}(2021)]%
        {bang2021estimating}
\bibfield{author}{\bibinfo{person}{Seungbae Bang}, \bibinfo{person}{Maria Korosteleva}, {and} \bibinfo{person}{Sung-Hee Lee}.} \bibinfo{year}{2021}\natexlab{}.
\newblock \showarticletitle{Estimating garment patterns from static scan data}. In \bibinfo{booktitle}{\emph{Computer Graphics Forum}}, Vol.~\bibinfo{volume}{40}. Wiley Online Library, \bibinfo{pages}{273--287}.
\newblock


\bibitem[Baraff and Witkin(1998)]%
        {BWCloth}
\bibfield{author}{\bibinfo{person}{David Baraff} {and} \bibinfo{person}{Andrew Witkin}.} \bibinfo{year}{1998}\natexlab{}.
\newblock \showarticletitle{Large steps in cloth simulation}. In \bibinfo{booktitle}{\emph{Proceedings of the 25th Annual Conference on Computer Graphics and Interactive Techniques}} \emph{(\bibinfo{series}{SIGGRAPH '98})}. \bibinfo{publisher}{Association for Computing Machinery}, \bibinfo{address}{New York, NY, USA}, \bibinfo{pages}{43–54}.
\newblock
\showISBNx{0897919998}
\urldef\tempurl%
\url{https://doi.org/10.1145/280814.280821}
\showDOI{\tempurl}


\bibitem[Bartle et~al\mbox{.}(2016)]%
        {bartle2016physics}
\bibfield{author}{\bibinfo{person}{Aric Bartle}, \bibinfo{person}{Alla Sheffer}, \bibinfo{person}{Vladimir~G Kim}, \bibinfo{person}{Danny~M Kaufman}, \bibinfo{person}{Nicholas Vining}, {and} \bibinfo{person}{Floraine Berthouzoz}.} \bibinfo{year}{2016}\natexlab{}.
\newblock \showarticletitle{Physics-driven pattern adjustment for direct 3D garment editing.}
\newblock \bibinfo{journal}{\emph{ACM Trans. Graph.}} \bibinfo{volume}{35}, \bibinfo{number}{4} (\bibinfo{year}{2016}), \bibinfo{pages}{50--1}.
\newblock


\bibitem[Bouaziz et~al\mbox{.}(2014)]%
        {bouaziz2014projective}
\bibfield{author}{\bibinfo{person}{Sofien Bouaziz}, \bibinfo{person}{Sebastian Martin}, \bibinfo{person}{Tiantian Liu}, \bibinfo{person}{Ladislav Kavan}, {and} \bibinfo{person}{Mark Pauly}.} \bibinfo{year}{2014}\natexlab{}.
\newblock \showarticletitle{Projective Dynamics: Fusing Constraint Projections for Fast Simulation}.
\newblock \bibinfo{journal}{\emph{Acm Transactions On Graphics}} \bibinfo{volume}{33}, \bibinfo{number}{4} (\bibinfo{year}{2014}), \bibinfo{pages}{154}.
\newblock


\bibitem[Brouet et~al\mbox{.}(2012)]%
        {brouet2012design}
\bibfield{author}{\bibinfo{person}{Remi Brouet}, \bibinfo{person}{Alla Sheffer}, \bibinfo{person}{Laurence Boissieux}, {and} \bibinfo{person}{Marie-Paule Cani}.} \bibinfo{year}{2012}\natexlab{}.
\newblock \showarticletitle{Design preserving garment transfer}.
\newblock \bibinfo{journal}{\emph{ACM Transactions on Graphics}} \bibinfo{volume}{31}, \bibinfo{number}{4} (\bibinfo{year}{2012}), \bibinfo{pages}{Article--No}.
\newblock


\bibitem[Chen et~al\mbox{.}(2024a)]%
        {chen2024vertex}
\bibfield{author}{\bibinfo{person}{Anka~He Chen}, \bibinfo{person}{Ziheng Liu}, \bibinfo{person}{Yin Yang}, {and} \bibinfo{person}{Cem Yuksel}.} \bibinfo{year}{2024}\natexlab{a}.
\newblock \showarticletitle{Vertex Block Descent}.
\newblock \bibinfo{journal}{\emph{arXiv preprint arXiv:2403.06321}} (\bibinfo{year}{2024}).
\newblock


\bibitem[Chen et~al\mbox{.}(2024b)]%
        {chen2024panelformer}
\bibfield{author}{\bibinfo{person}{Cheng-Hsiu Chen}, \bibinfo{person}{Jheng-Wei Su}, \bibinfo{person}{Min-Chun Hu}, \bibinfo{person}{Chih-Yuan Yao}, {and} \bibinfo{person}{Hung-Kuo Chu}.} \bibinfo{year}{2024}\natexlab{b}.
\newblock \showarticletitle{Panelformer: Sewing Pattern Reconstruction From 2D Garment Images}. In \bibinfo{booktitle}{\emph{Proceedings of the IEEE/CVF Winter Conference on Applications of Computer Vision}}. \bibinfo{pages}{454--463}.
\newblock


\bibitem[Chen et~al\mbox{.}(2022a)]%
        {chen2022virtual}
\bibfield{author}{\bibinfo{person}{Hsiao-yu Chen}, \bibinfo{person}{Edith Tretschk}, \bibinfo{person}{Tuur Stuyck}, \bibinfo{person}{Petr Kadlecek}, \bibinfo{person}{Ladislav Kavan}, \bibinfo{person}{Etienne Vouga}, {and} \bibinfo{person}{Christoph Lassner}.} \bibinfo{year}{2022}\natexlab{a}.
\newblock \showarticletitle{Virtual elastic objects}. In \bibinfo{booktitle}{\emph{Proceedings of the IEEE/CVF Conference on Computer Vision and Pattern Recognition}}. \bibinfo{pages}{15827--15837}.
\newblock


\bibitem[Chen et~al\mbox{.}(2022b)]%
        {chen2022structure}
\bibfield{author}{\bibinfo{person}{Xipeng Chen}, \bibinfo{person}{Guangrun Wang}, \bibinfo{person}{Dizhong Zhu}, \bibinfo{person}{Xiaodan Liang}, \bibinfo{person}{Philip Torr}, {and} \bibinfo{person}{Liang Lin}.} \bibinfo{year}{2022}\natexlab{b}.
\newblock \showarticletitle{Structure-Preserving 3D Garment Modeling with Neural Sewing Machines}.
\newblock \bibinfo{journal}{\emph{Advances in Neural Information Processing Systems}}  \bibinfo{volume}{35} (\bibinfo{year}{2022}), \bibinfo{pages}{15147--15159}.
\newblock


\bibitem[Choi and Ko(2002)]%
        {choi2002stable}
\bibfield{author}{\bibinfo{person}{Kwang-Jin Choi} {and} \bibinfo{person}{Hyeong-Seok Ko}.} \bibinfo{year}{2002}\natexlab{}.
\newblock \showarticletitle{Stable but responsive cloth}.
\newblock \bibinfo{journal}{\emph{ACM Transactions on Graphics (TOG)}} \bibinfo{volume}{21}, \bibinfo{number}{3} (\bibinfo{year}{2002}), \bibinfo{pages}{604--611}.
\newblock


\bibitem[de~Goes et~al\mbox{.}(2020)]%
        {fernando}
\bibfield{author}{\bibinfo{person}{Fernando de Goes}, \bibinfo{person}{Donald Fong}, {and} \bibinfo{person}{Meredith O'Malley}.} \bibinfo{year}{2020}\natexlab{}.
\newblock \showarticletitle{Garment Refitting for Digital Characters}. In \bibinfo{booktitle}{\emph{ACM SIGGRAPH 2020 Talks}} (Virtual Event, USA) \emph{(\bibinfo{series}{SIGGRAPH '20})}. \bibinfo{publisher}{Association for Computing Machinery}, \bibinfo{address}{New York, NY, USA}, Article \bibinfo{articleno}{74}, \bibinfo{numpages}{2}~pages.
\newblock
\showISBNx{9781450379717}
\urldef\tempurl%
\url{https://doi.org/10.1145/3388767.3407348}
\showDOI{\tempurl}


\bibitem[Du et~al\mbox{.}(2021)]%
        {du2021diffpd}
\bibfield{author}{\bibinfo{person}{Tao Du}, \bibinfo{person}{Kui Wu}, \bibinfo{person}{Pingchuan Ma}, \bibinfo{person}{Sebastien Wah}, \bibinfo{person}{Andrew Spielberg}, \bibinfo{person}{Daniela Rus}, {and} \bibinfo{person}{Wojciech Matusik}.} \bibinfo{year}{2021}\natexlab{}.
\newblock \showarticletitle{Diffpd: Differentiable projective dynamics}.
\newblock \bibinfo{journal}{\emph{ACM Transactions on Graphics (TOG)}} \bibinfo{volume}{41}, \bibinfo{number}{2} (\bibinfo{year}{2021}), \bibinfo{pages}{1--21}.
\newblock


\bibitem[Etzmu\ss et~al\mbox{.}(2003)]%
        {fastCloth}
\bibfield{author}{\bibinfo{person}{Olaf Etzmu\ss}, \bibinfo{person}{Michael Keckeisen}, {and} \bibinfo{person}{Wolfgang Stra{\ss}er}.} \bibinfo{year}{2003}\natexlab{}.
\newblock \showarticletitle{A Fast Finite Element Solution for Cloth Modelling}. In \bibinfo{booktitle}{\emph{Proceedings of the 11th Pacific Conference on Computer Graphics and Applications}} \emph{(\bibinfo{series}{PG '03})}. \bibinfo{publisher}{IEEE Computer Society}, \bibinfo{address}{USA}, \bibinfo{pages}{244}.
\newblock
\showISBNx{0769520286}


\bibitem[Gast et~al\mbox{.}(2015)]%
        {gast2015optimization}
\bibfield{author}{\bibinfo{person}{Theodore~F Gast}, \bibinfo{person}{Craig Schroeder}, \bibinfo{person}{Alexey Stomakhin}, \bibinfo{person}{Chenfanfu Jiang}, {and} \bibinfo{person}{Joseph~M Teran}.} \bibinfo{year}{2015}\natexlab{}.
\newblock \showarticletitle{Optimization integrator for large time steps}.
\newblock \bibinfo{journal}{\emph{IEEE transactions on visualization and computer graphics}} \bibinfo{volume}{21}, \bibinfo{number}{10} (\bibinfo{year}{2015}), \bibinfo{pages}{1103--1115}.
\newblock


\bibitem[Guo et~al\mbox{.}(2018)]%
        {guo2018material}
\bibfield{author}{\bibinfo{person}{Qi Guo}, \bibinfo{person}{Xuchen Han}, \bibinfo{person}{Chuyuan Fu}, \bibinfo{person}{Theodore Gast}, \bibinfo{person}{Rasmus Tamstorf}, {and} \bibinfo{person}{Joseph Teran}.} \bibinfo{year}{2018}\natexlab{}.
\newblock \showarticletitle{A material point method for thin shells with frictional contact}.
\newblock \bibinfo{journal}{\emph{ACM Transactions on Graphics (TOG)}} \bibinfo{volume}{37}, \bibinfo{number}{4} (\bibinfo{year}{2018}), \bibinfo{pages}{1--15}.
\newblock


\bibitem[Halimi et~al\mbox{.}(2022)]%
        {halimi2022pattern}
\bibfield{author}{\bibinfo{person}{Oshri Halimi}, \bibinfo{person}{Tuur Stuyck}, \bibinfo{person}{Donglai Xiang}, \bibinfo{person}{Timur Bagautdinov}, \bibinfo{person}{He Wen}, \bibinfo{person}{Ron Kimmel}, \bibinfo{person}{Takaaki Shiratori}, \bibinfo{person}{Chenglei Wu}, \bibinfo{person}{Yaser Sheikh}, {and} \bibinfo{person}{Fabian Prada}.} \bibinfo{year}{2022}\natexlab{}.
\newblock \showarticletitle{Pattern-based cloth registration and sparse-view animation}.
\newblock \bibinfo{journal}{\emph{ACM Transactions on Graphics (TOG)}} \bibinfo{volume}{41}, \bibinfo{number}{6} (\bibinfo{year}{2022}), \bibinfo{pages}{1--17}.
\newblock


\bibitem[He et~al\mbox{.}(2024)]%
        {he2024dresscode}
\bibfield{author}{\bibinfo{person}{Kai He}, \bibinfo{person}{Kaixin Yao}, \bibinfo{person}{Qixuan Zhang}, \bibinfo{person}{Jingyi Yu}, \bibinfo{person}{Lingjie Liu}, {and} \bibinfo{person}{Lan Xu}.} \bibinfo{year}{2024}\natexlab{}.
\newblock \bibinfo{title}{DressCode: Autoregressively Sewing and Generating Garments from Text Guidance}.
\newblock
\newblock
\showeprint[arxiv]{2401.16465}~[cs.CV]


\bibitem[Korosteleva and Lee(2022)]%
        {korosteleva2022neuraltailor}
\bibfield{author}{\bibinfo{person}{Maria Korosteleva} {and} \bibinfo{person}{Sung-Hee Lee}.} \bibinfo{year}{2022}\natexlab{}.
\newblock \showarticletitle{Neuraltailor: Reconstructing sewing pattern structures from 3d point clouds of garments}.
\newblock \bibinfo{journal}{\emph{ACM Transactions on Graphics (TOG)}} \bibinfo{volume}{41}, \bibinfo{number}{4} (\bibinfo{year}{2022}), \bibinfo{pages}{1--16}.
\newblock


\bibitem[Korosteleva and Sorkine-Hornung(2023)]%
        {korosteleva2023garmentcode}
\bibfield{author}{\bibinfo{person}{Maria Korosteleva} {and} \bibinfo{person}{Olga Sorkine-Hornung}.} \bibinfo{year}{2023}\natexlab{}.
\newblock \showarticletitle{GarmentCode: Programming Parametric Sewing Patterns}.
\newblock \bibinfo{journal}{\emph{ACM Transactions on Graphics (TOG)}} \bibinfo{volume}{42}, \bibinfo{number}{6} (\bibinfo{year}{2023}), \bibinfo{pages}{1--15}.
\newblock


\bibitem[Larionov et~al\mbox{.}(2022)]%
        {larionov2022estimating}
\bibfield{author}{\bibinfo{person}{Egor Larionov}, \bibinfo{person}{Marie-Lena Eckert}, \bibinfo{person}{Katja Wolff}, {and} \bibinfo{person}{Tuur Stuyck}.} \bibinfo{year}{2022}\natexlab{}.
\newblock \showarticletitle{Estimating cloth elasticity parameters using position-based simulation of compliant constrained dynamics}.
\newblock \bibinfo{journal}{\emph{arXiv preprint arXiv:2212.08790}} (\bibinfo{year}{2022}).
\newblock


\bibitem[Li et~al\mbox{.}(2023)]%
        {li2023garment}
\bibfield{author}{\bibinfo{person}{Ren Li}, \bibinfo{person}{Corentin Dumery}, \bibinfo{person}{Beno{\^\i}t Guillard}, {and} \bibinfo{person}{Pascal Fua}.} \bibinfo{year}{2023}\natexlab{}.
\newblock \showarticletitle{Garment Recovery with Shape and Deformation Priors}.
\newblock \bibinfo{journal}{\emph{arXiv preprint arXiv:2311.10356}} (\bibinfo{year}{2023}).
\newblock


\bibitem[Li et~al\mbox{.}(2024b)]%
        {li2024isp}
\bibfield{author}{\bibinfo{person}{Ren Li}, \bibinfo{person}{Beno{\^\i}t Guillard}, {and} \bibinfo{person}{Pascal Fua}.} \bibinfo{year}{2024}\natexlab{b}.
\newblock \showarticletitle{Isp: Multi-layered garment draping with implicit sewing patterns}.
\newblock \bibinfo{journal}{\emph{Advances in Neural Information Processing Systems}}  \bibinfo{volume}{36} (\bibinfo{year}{2024}).
\newblock


\bibitem[Li et~al\mbox{.}(2024a)]%
        {li2023diffavatar}
\bibfield{author}{\bibinfo{person}{Yifei Li}, \bibinfo{person}{Hsiao-yu Chen}, \bibinfo{person}{Egor Larionov}, \bibinfo{person}{Nikolaos Sarafianos}, \bibinfo{person}{Wojciech Matusik}, {and} \bibinfo{person}{Tuur Stuyck}.} \bibinfo{year}{2024}\natexlab{a}.
\newblock \showarticletitle{DiffAvatar: Simulation-Ready Garment Optimization with Differentiable Simulation}.
\newblock \bibinfo{journal}{\emph{IEEE Conference on Computer Vision and Pattern Recognition(CVPR)}} (\bibinfo{year}{2024}).
\newblock


\bibitem[Li et~al\mbox{.}(2022)]%
        {li2022diffcloth}
\bibfield{author}{\bibinfo{person}{Yifei Li}, \bibinfo{person}{Tao Du}, \bibinfo{person}{Kui Wu}, \bibinfo{person}{Jie Xu}, {and} \bibinfo{person}{Wojciech Matusik}.} \bibinfo{year}{2022}\natexlab{}.
\newblock \showarticletitle{Diffcloth: Differentiable cloth simulation with dry frictional contact}.
\newblock \bibinfo{journal}{\emph{ACM Transactions on Graphics (TOG)}} \bibinfo{volume}{42}, \bibinfo{number}{1} (\bibinfo{year}{2022}), \bibinfo{pages}{1--20}.
\newblock


\bibitem[Liang et~al\mbox{.}(2019)]%
        {liang2019differentiable}
\bibfield{author}{\bibinfo{person}{Junbang Liang}, \bibinfo{person}{Ming Lin}, {and} \bibinfo{person}{Vladlen Koltun}.} \bibinfo{year}{2019}\natexlab{}.
\newblock \showarticletitle{Differentiable cloth simulation for inverse problems}.
\newblock \bibinfo{journal}{\emph{Advances in Neural Information Processing Systems}}  \bibinfo{volume}{32} (\bibinfo{year}{2019}).
\newblock


\bibitem[Lipman et~al\mbox{.}(2008)]%
        {lipman2008green}
\bibfield{author}{\bibinfo{person}{Yaron Lipman}, \bibinfo{person}{David Levin}, {and} \bibinfo{person}{Daniel Cohen-Or}.} \bibinfo{year}{2008}\natexlab{}.
\newblock \showarticletitle{Green coordinates}.
\newblock \bibinfo{journal}{\emph{ACM transactions on graphics (TOG)}} \bibinfo{volume}{27}, \bibinfo{number}{3} (\bibinfo{year}{2008}), \bibinfo{pages}{1--10}.
\newblock


\bibitem[Liu et~al\mbox{.}(2023)]%
        {liu2023towards}
\bibfield{author}{\bibinfo{person}{Lijuan Liu}, \bibinfo{person}{Xiangyu Xu}, \bibinfo{person}{Zhijie Lin}, \bibinfo{person}{Jiabin Liang}, {and} \bibinfo{person}{Shuicheng Yan}.} \bibinfo{year}{2023}\natexlab{}.
\newblock \showarticletitle{Towards garment sewing pattern reconstruction from a single image}.
\newblock \bibinfo{journal}{\emph{ACM Transactions on Graphics (TOG)}} \bibinfo{volume}{42}, \bibinfo{number}{6} (\bibinfo{year}{2023}), \bibinfo{pages}{1--15}.
\newblock


\bibitem[Loper et~al\mbox{.}(2015)]%
        {SMPL}
\bibfield{author}{\bibinfo{person}{Matthew Loper}, \bibinfo{person}{Naureen Mahmood}, \bibinfo{person}{Javier Romero}, \bibinfo{person}{Gerard Pons-Moll}, {and} \bibinfo{person}{Michael~J. Black}.} \bibinfo{year}{2015}\natexlab{}.
\newblock \showarticletitle{{SMPL}: A Skinned Multi-Person Linear Model}.
\newblock \bibinfo{journal}{\emph{ACM Trans. Graphics (Proc. SIGGRAPH Asia)}} \bibinfo{volume}{34}, \bibinfo{number}{6} (\bibinfo{date}{Oct.} \bibinfo{year}{2015}), \bibinfo{pages}{248:1--248:16}.
\newblock


\bibitem[Macklin et~al\mbox{.}(2016)]%
        {macklin2016xpbd}
\bibfield{author}{\bibinfo{person}{Miles Macklin}, \bibinfo{person}{Matthias M{\"u}ller}, {and} \bibinfo{person}{Nuttapong Chentanez}.} \bibinfo{year}{2016}\natexlab{}.
\newblock \showarticletitle{XPBD: position-based simulation of compliant constrained dynamics}. In \bibinfo{booktitle}{\emph{Proceedings of the 9th International Conference on Motion in Games}}. \bibinfo{pages}{49--54}.
\newblock


\bibitem[Martin et~al\mbox{.}(2011)]%
        {martin2011example}
\bibfield{author}{\bibinfo{person}{Sebastian Martin}, \bibinfo{person}{Bernhard Thomaszewski}, \bibinfo{person}{Eitan Grinspun}, {and} \bibinfo{person}{Markus Gross}.} \bibinfo{year}{2011}\natexlab{}.
\newblock \showarticletitle{Example-based elastic materials}.
\newblock In \bibinfo{booktitle}{\emph{ACM SIGGRAPH 2011 papers}}. \bibinfo{pages}{1--8}.
\newblock


\bibitem[M{\"u}ller et~al\mbox{.}(2007)]%
        {muller2007position}
\bibfield{author}{\bibinfo{person}{Matthias M{\"u}ller}, \bibinfo{person}{Bruno Heidelberger}, \bibinfo{person}{Marcus Hennix}, {and} \bibinfo{person}{John Ratcliff}.} \bibinfo{year}{2007}\natexlab{}.
\newblock \showarticletitle{Position based dynamics}.
\newblock \bibinfo{journal}{\emph{Journal of Visual Communication and Image Representation}} \bibinfo{volume}{18}, \bibinfo{number}{2} (\bibinfo{year}{2007}), \bibinfo{pages}{109--118}.
\newblock


\bibitem[Pietroni et~al\mbox{.}(2022)]%
        {pietroni2022computational}
\bibfield{author}{\bibinfo{person}{Nico Pietroni}, \bibinfo{person}{Corentin Dumery}, \bibinfo{person}{Raphael Falque}, \bibinfo{person}{Mark Liu}, \bibinfo{person}{Teresa~A Vidal-Calleja}, {and} \bibinfo{person}{Olga Sorkine-Hornung}.} \bibinfo{year}{2022}\natexlab{}.
\newblock \showarticletitle{Computational pattern making from 3D garment models.}
\newblock \bibinfo{journal}{\emph{ACM Trans. Graph.}} \bibinfo{volume}{41}, \bibinfo{number}{4} (\bibinfo{year}{2022}), \bibinfo{pages}{157--1}.
\newblock


\bibitem[Sarafianos et~al\mbox{.}(2024)]%
        {sarafianos2024garment3dgen}
\bibfield{author}{\bibinfo{person}{Nikolaos Sarafianos}, \bibinfo{person}{Tuur Stuyck}, \bibinfo{person}{Xiaoyu Xiang}, \bibinfo{person}{Yilei Li}, \bibinfo{person}{Jovan Popovic}, {and} \bibinfo{person}{Rakesh Ranjan}.} \bibinfo{year}{2024}\natexlab{}.
\newblock \showarticletitle{Garment3DGen: 3D Garment Stylization and Texture Generation}.
\newblock \bibinfo{journal}{\emph{arXiv preprint arXiv:2403.18816}} (\bibinfo{year}{2024}).
\newblock


\bibitem[Shewchuk(2002)]%
        {shewchuk2002quality}
\bibfield{author}{\bibinfo{person}{Jonathan~Richard Shewchuk}.} \bibinfo{year}{2002}\natexlab{}.
\newblock \showarticletitle{What is a Good Linear Element? Interpolation, Conditioning, and Quality Measures}. In \bibinfo{booktitle}{\emph{International Meshing Roundtable Conference}}.
\newblock
\urldef\tempurl%
\url{https://api.semanticscholar.org/CorpusID:8691914}
\showURL{%
\tempurl}


\bibitem[Stuyck(2022)]%
        {stuyck2022cloth}
\bibfield{author}{\bibinfo{person}{Tuur Stuyck}.} \bibinfo{year}{2022}\natexlab{}.
\newblock \bibinfo{booktitle}{\emph{Cloth simulation for computer graphics}}.
\newblock \bibinfo{publisher}{Springer Nature}.
\newblock


\bibitem[Stuyck and Chen(2023)]%
        {stuyck2023diffxpbd}
\bibfield{author}{\bibinfo{person}{Tuur Stuyck} {and} \bibinfo{person}{Hsiao-yu Chen}.} \bibinfo{year}{2023}\natexlab{}.
\newblock \showarticletitle{Diffxpbd: Differentiable position-based simulation of compliant constraint dynamics}.
\newblock \bibinfo{journal}{\emph{Proceedings of the ACM on Computer Graphics and Interactive Techniques}} \bibinfo{volume}{6}, \bibinfo{number}{3} (\bibinfo{year}{2023}), \bibinfo{pages}{1--14}.
\newblock


\bibitem[Terzopoulos et~al\mbox{.}(1987)]%
        {edm}
\bibfield{author}{\bibinfo{person}{Demetri Terzopoulos}, \bibinfo{person}{John Platt}, \bibinfo{person}{Alan Barr}, {and} \bibinfo{person}{Kurt Fleischer}.} \bibinfo{year}{1987}\natexlab{}.
\newblock \showarticletitle{Elastically deformable models}.
\newblock \bibinfo{journal}{\emph{SIGGRAPH Comput. Graph.}} \bibinfo{volume}{21}, \bibinfo{number}{4} (\bibinfo{date}{aug} \bibinfo{year}{1987}), \bibinfo{pages}{205–214}.
\newblock
\showISSN{0097-8930}
\urldef\tempurl%
\url{https://doi.org/10.1145/37402.37427}
\showDOI{\tempurl}


\bibitem[Wang(2018)]%
        {wang2018rule}
\bibfield{author}{\bibinfo{person}{Huamin Wang}.} \bibinfo{year}{2018}\natexlab{}.
\newblock \showarticletitle{Rule-free sewing pattern adjustment with precision and efficiency}.
\newblock \bibinfo{journal}{\emph{ACM Transactions on Graphics (TOG)}} \bibinfo{volume}{37}, \bibinfo{number}{4} (\bibinfo{year}{2018}), \bibinfo{pages}{1--13}.
\newblock


\bibitem[Wojtan et~al\mbox{.}(2006)]%
        {wojtan2006keyframe}
\bibfield{author}{\bibinfo{person}{Chris Wojtan}, \bibinfo{person}{Peter~J Mucha}, {and} \bibinfo{person}{Greg Turk}.} \bibinfo{year}{2006}\natexlab{}.
\newblock \showarticletitle{Keyframe control of complex particle systems using the adjoint method}. In \bibinfo{booktitle}{\emph{Proceedings of the 2006 ACM SIGGRAPH/Eurographics symposium on Computer animation}}. \bibinfo{pages}{15--23}.
\newblock


\bibitem[Wolff et~al\mbox{.}(2023)]%
        {wolff2023designing}
\bibfield{author}{\bibinfo{person}{Katja Wolff}, \bibinfo{person}{Philipp Herholz}, \bibinfo{person}{Verena Ziegler}, \bibinfo{person}{Frauke Link}, \bibinfo{person}{Nico Br{\"u}gel}, {and} \bibinfo{person}{Olga Sorkine-Hornung}.} \bibinfo{year}{2023}\natexlab{}.
\newblock \showarticletitle{Designing Personalized Garments with Body Movement}. In \bibinfo{booktitle}{\emph{Computer Graphics Forum}}, Vol.~\bibinfo{volume}{42}. Wiley Online Library, \bibinfo{pages}{180--194}.
\newblock


\bibitem[Xiang et~al\mbox{.}(2022)]%
        {xiang2022dressing}
\bibfield{author}{\bibinfo{person}{Donglai Xiang}, \bibinfo{person}{Timur Bagautdinov}, \bibinfo{person}{Tuur Stuyck}, \bibinfo{person}{Fabian Prada}, \bibinfo{person}{Javier Romero}, \bibinfo{person}{Weipeng Xu}, \bibinfo{person}{Shunsuke Saito}, \bibinfo{person}{Jingfan Guo}, \bibinfo{person}{Breannan Smith}, \bibinfo{person}{Takaaki Shiratori}, {et~al\mbox{.}}} \bibinfo{year}{2022}\natexlab{}.
\newblock \showarticletitle{Dressing avatars: Deep photorealistic appearance for physically simulated clothing}.
\newblock \bibinfo{journal}{\emph{ACM Transactions on Graphics (TOG)}} \bibinfo{volume}{41}, \bibinfo{number}{6} (\bibinfo{year}{2022}), \bibinfo{pages}{1--15}.
\newblock


\bibitem[Yang et~al\mbox{.}(2018)]%
        {garmentFromImage}
\bibfield{author}{\bibinfo{person}{Shan Yang}, \bibinfo{person}{Zherong Pan}, \bibinfo{person}{Tanya Amert}, \bibinfo{person}{Ke Wang}, \bibinfo{person}{Licheng Yu}, \bibinfo{person}{Tamara Berg}, {and} \bibinfo{person}{Ming~C. Lin}.} \bibinfo{year}{2018}\natexlab{}.
\newblock \showarticletitle{Physics-Inspired Garment Recovery from a Single-View Image}.
\newblock \bibinfo{journal}{\emph{ACM Trans. Graph.}} \bibinfo{volume}{37}, \bibinfo{number}{5}, Article \bibinfo{articleno}{170} (\bibinfo{date}{nov} \bibinfo{year}{2018}), \bibinfo{numpages}{14}~pages.
\newblock
\showISSN{0730-0301}
\urldef\tempurl%
\url{https://doi.org/10.1145/3026479}
\showDOI{\tempurl}


\bibitem[Zheng et~al\mbox{.}(2024)]%
        {PhysAavatar24}
\bibfield{author}{\bibinfo{person}{Yang Zheng}, \bibinfo{person}{Qingqing Zhao}, \bibinfo{person}{Guandao Yang}, \bibinfo{person}{Wang Yifan}, \bibinfo{person}{Donglai Xiang}, \bibinfo{person}{Florian Dubost}, \bibinfo{person}{Dmitry Lagun}, \bibinfo{person}{Thabo Beeler}, \bibinfo{person}{Federico Tombari}, \bibinfo{person}{Leonidas Guibas}, {and} \bibinfo{person}{Gordon Wetzstein}.} \bibinfo{year}{2024}\natexlab{}.
\newblock \showarticletitle{PhysAvatar: Learning the Physics of Dressed 3D Avatars from Visual Observations}.
\newblock \bibinfo{journal}{\emph{arxiv}}.
\newblock


\end{thebibliography}

\end{document}